\def\Msun{\hbox{M$_\odot$}}

\def\cm3{\hbox{cm$^{-3}$}}
\def\s1{\hbox{s$^{-1}$}}

\def\one{\,{\sc i}}             
\def\two{\,{\sc ii}}
\def\three{\,{\sc iii}}
\def\four{\,{\sc iv}}
\def\five{\,{\sc v}}
\def\six{\,\textsc{vi}}

\def\ha{H$\alpha$}
\def\pa{P$\alpha$}
\def\hst{{\it HST \/}}

\documentclass[twocolumn]{aastex63}

\usepackage{amsmath}
\usepackage{amssymb}
\usepackage{mathrsfs}
\usepackage{upgreek}
\usepackage{graphicx}
\usepackage{overpic}
\usepackage{threeparttable}
\received{2020 March 20}
\revised{2020 April 13}
\accepted{\today}
\submitjournal{ApJ}

\shorttitle{The Three Nuclear Super Star Clusters in  NGC 5253}
\shortauthors{L.\ J.\ Smith et al.}

\begin{document}


\title{The Three Young Nuclear Super Star Clusters in NGC 5253}
\correspondingauthor{Linda\ J.\ Smith}
\email{lsmith@stsci.edu}
\author[0000-0002-0806-168X]{Linda J. Smith}
\affiliation{Space Telescope Science Institute and European Space Agency, 3700 San Martin Drive, Baltimore, MD 21218}
\author{Varun Bajaj}
\affiliation{Space Telescope Science Institute, 3700 San Martin Drive, Baltimore, MD 21218}
\author{Jenna Ryon}
\affiliation{Space Telescope Science Institute, 3700 San Martin Drive, Baltimore, MD 21218}
\author[0000-0003-2954-7643]{Elena Sabbi}
\affiliation{Space Telescope Science Institute, 3700 San Martin Drive, Baltimore, MD 21218}

\begin{abstract}
The blue compact dwarf galaxy NGC~5253 hosts a very young central starburst. The center contains intense radio thermal emission from a massive ultracompact H\two\ region (or ``supernebula'') and two massive and very young super star clusters (SSCs), which are seen at optical and infrared wavelengths. The spatial correspondence between these three objects over an area of $< 0\arcsec.5$ is very uncertain, and it is usually assumed that one of the optically-detected SSCs coincides with the supernebula. Using the Gaia Data Release 2 catalog, we improve the absolute astrometry of  {\it Hubble Space Telescope\/} ultraviolet, optical and infrared images by a factor of $\sim 10$ and match them to the radio observations with an accuracy of 10--20 mas. 
 We find that there are 3 SSCs at the center of NGC~5253. At optical wavelengths, the two SSCs lie either side of the supernebula, which coincides with a highly obscured region. In the infrared, all 3 SSCs are seen with the supernebula dominating at the longest wavelengths. 
We examine the properties of the SSCs, which have ages of $\sim 1$ Myr, are very compact ($<0.6$~pc) and separated by projected distances of only 1.5 and 5.8~pc. It seems likely that they will merge and may form the building blocks for nuclear star clusters.
\end{abstract}

\keywords{galaxies: dwarf -- galaxies: individual (NGC\,5253) -- galaxies: starburst -- galaxies: star clusters}

\section{Introduction} \label{intro}

A high pressure environment ($P/k_{\rm B} >10^6\,{\rm cm}^{-3}\,{\rm K}$) is thought to be a prerequisite for both the formation of globular clusters (GCs) in redshift $> 2$ galaxies and young 
super star clusters (SSCs\footnote{defined as clusters with masses $> 10^5$~\Msun}) in the nearby universe \citep{elmegreen97, 
kravtsov05, kruijssen15, elmegreen18}. High gas pressures are naturally found in the turbulent, clumpy, gas-rich disks of young galaxies \citep{forster09, forster18, elmegreen05, swinbank11}, 
or in dense infalling gas streams from cold accretion \citep{mandelker18}. High pressures are rare in the quiescent disks of nearby galaxies \citep[e.g.][]{ Kruijssen13}. 
Instead, nearby examples of SSCs are usually confined to galaxy mergers, starburst dwarf galaxies and nuclear regions of galaxies where pressures are high \citep*{portegies10}.

The blue compact dwarf galaxy NGC~5253 is a good example of SSC formation in nearby galaxies. It has a metallicity of $12+\log\,$O/H$=8.26$ or
37\% solar \citep*{monreal12} (using the solar oxygen abundance of $12+\log\,$O/H$=8.69$; \citet{asplund09}) and hosts a very young central starburst. 
There is a rich population of SSCs in the central starburst and surrounding 300~pc region \citep{degrijs13, calzetti15}. The current starburst is generally accepted to be 
triggered by infalling material along the minor axis, detected in CO(2-1) \citep{meier02, miura18} and H\one\ \citep{kobulnicky08,lopez12}. 
External gas infall has been investigated by \citet{verbeke14} and their simulations show that an infall  can trigger a starburst in blue compact dwarf galaxies.

The center of NGC~5253 has been extensively studied from radio to X-ray wavelengths. \citet{beck96} discovered intense radio thermal continuum emission from a massive ultracompact H\two\ region ionized by a young SSC. This radio source was resolved and termed the ``supernebula'' \citep*{turner00,turner04}. In the infrared (IR), \citet{alonso04} identified a double nuclear star cluster from {\it HST}/NICMOS observations. The properties of these two clusters were investigated by \citet{calzetti15} using photometry from 13 \hst bands covering the far-ultraviolet (FUV) to the IR. 

Observations of the molecular content of NGC~5253 show that CO emission is mainly confined to the infalling streamer along the minor axis and the central starburst region \citep{meier02, miura15, turner15, turner17, miura18}. Unusually high star formation efficiencies (SFEs) of 35--80\% have been found by these authors for the central region.

The spatial correspondence between the radio ``supernebula'' and the two optical/IR clusters is very uncertain because of the need for exquisite absolute astrometry over an area of $< 0\arcsec.5$. The absolute astrometry of \hst is limited by the accuracy of Guide Star Catalog II \citep{lasker08} to $0\arcsec.1$--$0\arcsec.3$ \citep{koekemoer06}. With the publication of Gaia Data Release 2 (DR2) \citep{gaia16,dr2}, it is now possible to match the \hst imaging with the radio observations to reveal the cluster content of the center of NGC~5253.

In this paper, we re-map \hst images of the center of NGC~5253 to the Gaia DR2 reference frame and successfully match the radio to the optical and IR observations with an accuracy of $< 20$~mas. 
In Section~\ref{views}, we describe the radio, IR and optical views of the center. In Section~\ref{astrom}, the re-mapping of the \hst data is described and the absolute astrometry is compared with the radio data. We also examine the size of the clusters from the \hst imaging.
In Section~\ref{disc}, we discuss the properties of the clusters, and present our conclusions.

Distances in the literature for NGC 5253 range from 3--4~Mpc. We adopt the most recent distance measurement to NGC~5253 of $3.32\pm0.25$ Mpc determined from the tip of the red giant branch \citep{sabbi18}. At this distance, $0\arcsec.1 = 1.6$ pc. 

\section{The center of NGC 5253 as seen at different wavelengths}\label{views}
\subsection{The radio view}
The deeply embedded ultracompact H\two\ region \citep{beck96}  at the center of NGC 5253 has been extensively studied with the Very Large Array (VLA) and the Atacama Large Millimeter/Submillimeter Array (ALMA). \citet{turner00} observed this source at 1.3 and 2~cm with the VLA and found one dominant radio source, which they termed the ``supernebula'', and a secondary source $0\arcsec.23$ to the East  detected only at 1.3~cm. The supernebula was resolved by \citet{turner04} at 7~mm to have a bright core with a FWHM size of $99\pm9 \times 39\pm4$~mas or $1.59 \times 0.63$~pc for our adopted distance. They derive a density for the bright core of 3--4$\times 10^4$~\cm3 and a flux of ionizing Lyman continuum photons for the bright core of $1.4\times10^{52}$~\s1. The coordinates of the supernebula core and secondary source are given in Table~\ref{tab:coords}.  

\citet{turner17} and \citet{consiglio17}  present ALMA CO(3-2) observations of NGC 5253. They find a number of dense clouds within the central $\sim 100$~pc  starburst region, previously identified as ``Cloud D'' by \citet{meier02}. \citet{turner17} focus on the compact ``Cloud D1'' which is very nearly coincident with the supernebula (see coordinates in Table~\ref{tab:coords}; offset is 35~mas or 0.56~pc in projection). They derive a size for D1 of $220\pm33 \times 100\pm54$ mas or $3.5 \times 1.6$~pc and find a virial mass from the CO line width of $\sim 2.5 \pm 0.9 \times 10^5$~\Msun. They suggest that Cloud D1 consists of hot molecular clumps or cores associated with the stars in the embedded supernebula cluster. They associate D1 with the supernebula on the basis of their close spatial and velocity coincidences. They also find that D1 is optically thin in CO(3-2) suggesting it is hot.

\citet{miura18} present deep, high resolution  ALMA CO(2-1) observations of NGC~5253 and identify 118 molecular clouds. Although they reserve the details of the starburst region to a future paper, they note that the clouds near the central starburst have large velocity widths, high gas surface densities, and high thermal pressures of $P/k_{\rm B} \ga 10^6$--$10^7$~\cm3K. 

\citet{bendo17} have observed H30$\alpha$ emission with ALMA  and find a very bright central peak with a de-convolved angular size of $\sim 0\arcsec.15$ or 2.4~pc. 

\begin{deluxetable*}{lllllcl}
\tablecaption{Coordinates of central sources in NGC~5253
\label{tab:coords}}
\tablecolumns{7}
\tablewidth{0pt}
\tablehead{
\colhead{Reference}  & \colhead{Designation} &
\colhead{RA} & \colhead{Dec}  & \colhead{Accuracy} & \colhead{Telescope} &\colhead{Wavelength}\\
&&(J2000)&(J2000)&\colhead{(mas)}
}
\startdata
Turner et al. (2000) & Supernebula & 13\ 39\ 55.964 & $-31$\ 38\ 24.38 & $\pm 10$& VLA & 2, 1.3 cm\\
& Secondary-East & 13\ 39\ 55.982\ & $-31$\ 38\ 24.37 & $\pm 10$ & VLA &1.3 cm\\
Turner \& Beck (2004) & Supernebula Core & 13\ 39\ 55.9631 & $-31$\ 38\ 24.388 & $\pm 4$& VLA, Pie Town & 7 mm\\
Turner et al. (2017) & Cloud D1 & 13\ 39\ 55.9651 & $-31$\ 38\ 24.364 & $\pm 6$ & ALMA & CO(3-2) \\
Calzetti et. al. (2015) & Cluster 5 & 13\ 39\ 55.986 & $-31$\ 38\  24.54 & 100--300 & HST/HRC & \ha\\
& Cluster 11 & 13\ 39\ 55.951 & $-31$\ 38\ 24.45 &100--300 & HST/NICMOS & \pa\\
This paper & Cluster 5 & 13\ 39\ 55.9914 & $-31$\ 38\ 24.399 & $\pm 12$ & HST/HRC & F814W\\
& Cluster 11 & 13\ 39\ 55.9568 & $-31$\ 38\ 24.339 & $\pm12$ & HST/HRC & F814W\\
\enddata
\end{deluxetable*}

\subsection{The infrared view}\label{ir}
\citet{alonso04} presented {\it HST}/NICMOS observations of the central region of NGC 5253 and discovered a double nuclear star cluster separated by $0\arcsec.3$--$0\arcsec.4$ that appears to be coincident with the primary and secondary radio sources of \citet{turner00}.
\citet{turner04} also compare their 7~mm radio image with the NICMOS data and suggest that the western cluster is probably coincident with the supernebula, although they note that the absolute pointing accuracy of  \hst ($\sim 1\arcsec$ at the epoch of the observations) is inadequate to register the two sets of images. 

In a recent paper, \citet{cohen18} present observations of Br$\alpha$ $4.05\mu$m emission across the supernebula and K-band imaging from the slit-viewing camera at a resolution of $0\arcsec.1$ taken with NIRSPEC on the Keck II telescope in adaptive optics mode.  They investigate the relative positions of the two optical/IR clusters and the supernebula by aligning clusters in the field of view of the K-band image with the \hst F814W image and the VLT and ALMA radio images  \citep{turner04, turner17}. At the longer wavelengths, they assume that the supernebula is responsible for the free-free emission and the K-band peak. These relative alignments (accurate to $\pm50$ mas) show that neither of the two optical/IR clusters are coincident with the supernebula, but are offset by 0\arcsec.35 and  0\arcsec.14. 

\subsection{The optical view}\label{opt}
\citet{calzetti15} present an  {\it HST}-based study of the brightest young star clusters in NGC 5253. They derive the properties of the two nuclear clusters (\#5 and \#11 in their terminology) by fitting spectral energy distributions (SEDs) using photometry from 13  \hst filters covering the far-UV to the IR. They find that the two clusters are extremely young with ages of $1\pm1$~Myr and have masses of $7.5\pm0.3 \times 10^4$ and $2.5 \pm0.6\times 10^5$~\Msun\ for \#5 and \#11 respectively. The western cluster \#11 is heavily reddened and a combination of a homogeneous dust-star mixture and a foreground dust screen is required for the SED fit, while the cluster \#5 SED can be fit with a foreground screen. \citet{calzetti15} define the positions of clusters \#5 and \#11 as corresponding to the peaks of the \ha\ and \pa\ emission respectively. They find that the two clusters are close to the primary and secondary radio sources detected by \citet{turner00} and probably coincident, given the uncertainty of \hst absolute astrometry (Table~\ref{tab:coords}).

\citet{smith16} examined \hst ultraviolet and Very Large Telescope optical spectroscopy of cluster \#5 and confirmed the young age of $1-2$~Myr, and showed that very massive stars ($> 100 \Msun$) must be present to explain the stellar features and high ionizing flux.

\begin{deluxetable*}{lclcll}
\tablecolumns{6}
\tablecaption{HST Archive Images of NGC~5253\label{tab:images}}
\tablewidth{0pt}
\tablehead{
\colhead{Instrument/} & \colhead{Drizzled Pixel Size}  & \colhead{Filter}  &  \colhead{Exposure Time}
& \colhead{Date of} & \colhead{GO}\\
\colhead{Camera} &\colhead{(arcsec)}  
& &\colhead{(s)} & {Observation} & \colhead{Program}
}
\startdata
ACS/WFC     & 0.04 & F814W & 2360 & 2005 Dec 27 & 10765 \\
ACS/SBC     & 0.025 & F125LP & 2660  &  2009 Mar 07 &11579 \\
ACS/HRC     & 0.025 &  F330W & 1796  &  2006 Feb 20 &10609 \\
            &       & F435W & 600 &  2006 Feb 20 &10609 \\
            &       & F550M & 800 &  2006 Feb 20 & 10609 \\
            &       & F658N & 240 &  2006 Feb 20& 10609 \\ 
            &       & F814W & 368 &  2006 Feb 20 & 10609 \\ 
NICMOS/NIC2 & 0.04  & F110W & 96& 1998 Jan 04 & 7219\\ 
            &       & F160W & 96 & 1998  Jan 04 &7219\\ 
            &        & F187N & 256 & 1998 Jan 04&7219\\ 
            &       & F190N &256 & 1998 Jan 04&7219\\ 
            &        & F222M & 640 & 1998 Jan 04&7219\\ 
\enddata
\end{deluxetable*}

\section{The multi-wavelength view of the center of NGC 5253}\label{astrom}
\subsection{Remapping of the \hst images}

We now consider the spatial correspondence between the clusters identified at radio, IR and optical wavelengths by re-mapping them to the Gaia reference frame.

In Table~\ref{tab:images}, the \hst images are listed for which we have remapped the astrometry to the Gaia DR2 \citep{gaia16,dr2} reference frame. The 
registration was performed using the DrizzlePac\footnote{http://www.stsci.edu/scientific-community/software/drizzlepac}
 function \verb+tweakreg+, which aligns sources found in an input HST image (ACS/WFC F814W) to positions from a reference catalog (Gaia DR2). The ACS/WFC F814W image was used as input because of its larger field of view compared to ACS/HRC. 
 
 Proper motions were applied to the source positions to correct for the difference in epoch between Gaia and the ACS F814W observations (9.5 yr).  
 We filtered out sources giving poor residuals in the overall alignment by considering the normalized proper motions (proper motion divided by the standard uncertainty in proper motion as listed in the Gaia DR2 catalog). We empirically determined the best threshold for the normalized proper motions as $\ge 0.5$ to minimize the residuals given by  \verb+tweakreg+ over the whole WFC F814W image.
 The World Coordinate System (WCS) of the F814W image was then recalculated using the transforms derived from the alignment. With 65 Gaia sources within 4 arcmin of the center of NGC~5253, we find that the alignment accuracy is $\pm10$~mas from the residuals. The ACS/HRC and NICMOS/NIC2 images were then aligned to the transformed ACS/WFC F814W image (and hence Gaia DR2) using as many sources in the images as possible. For ACS/HRC, we used 150 sources on average, and the alignment is accurate to $\pm12$~ mas. The NICMOS/NIC2 images were aligned using 100 sources on average, and the alignment is $\pm 20$~ mas.
 
\begin{figure*}
\gridline{\fig{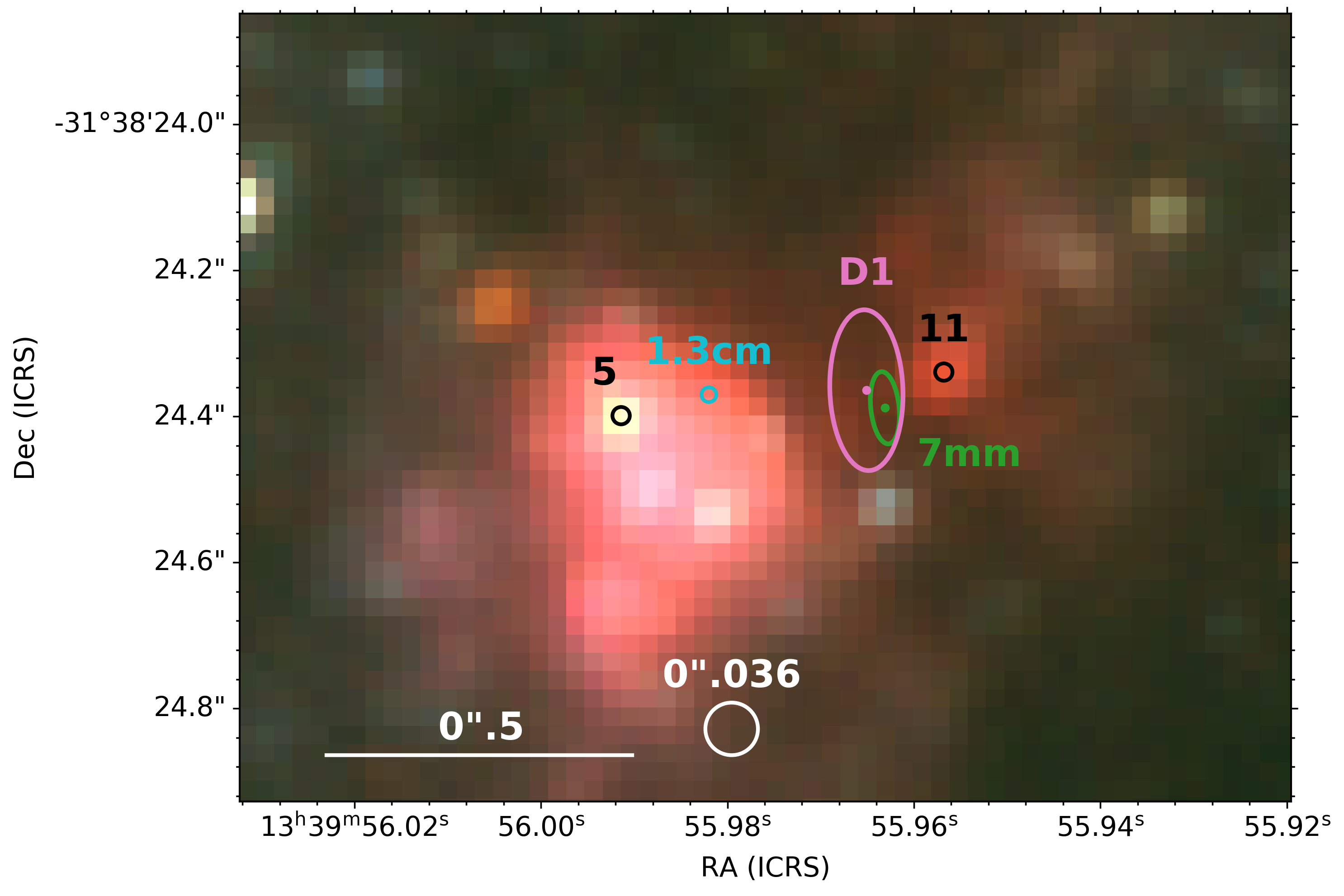}{0.47\textwidth}{(a)}
\fig{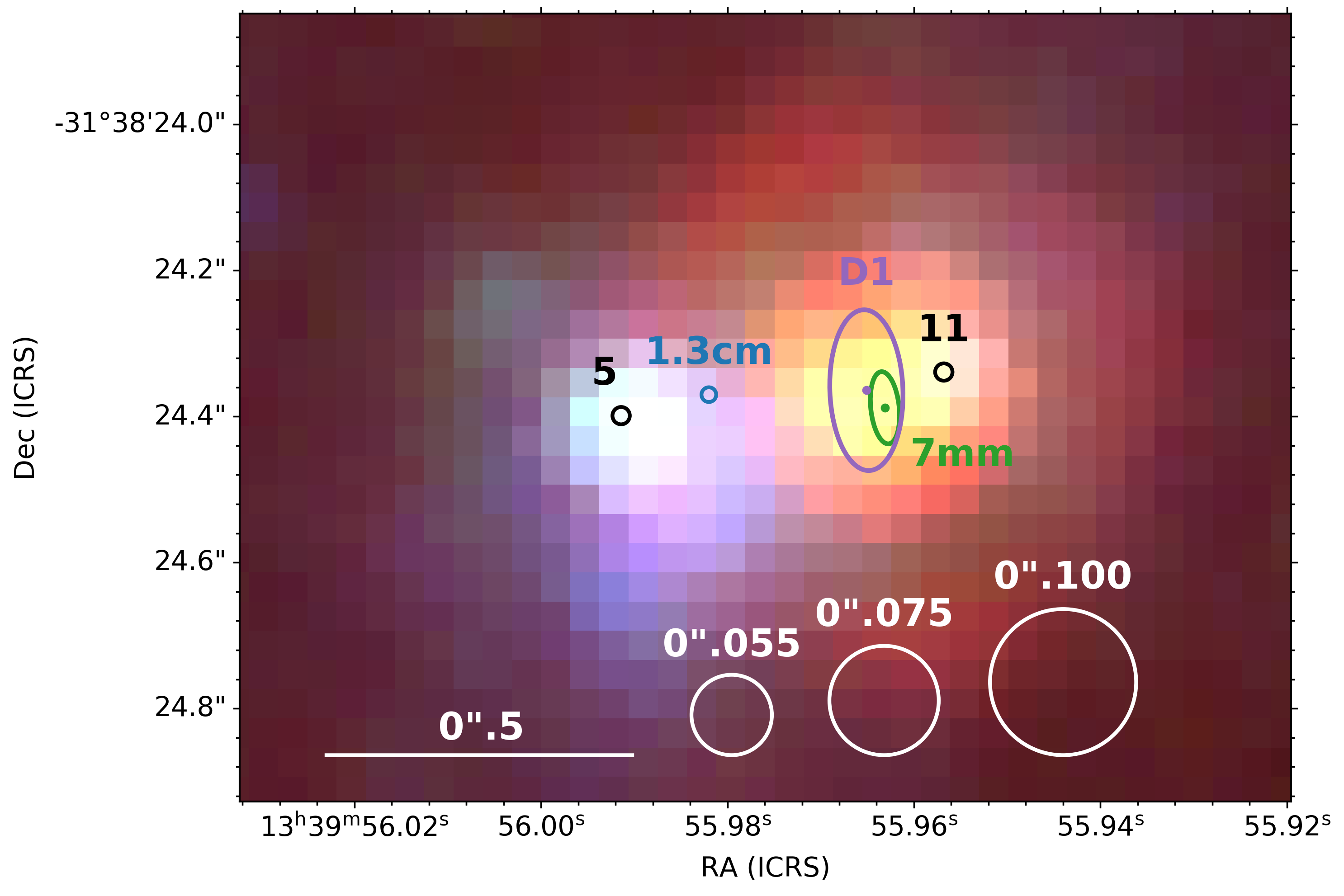}{0.47\textwidth}{(b)}}
\caption{(a) ACS/HRC F435W, F550M, F814W image and (b) NICMOS/NIC2 F110W F160W, F222M image of the central region of NGC 5253.
The NICMOS image is oversampled with a pixel size of 0\arcsec.04. The optical/IR clusters \#5 and \#11 are indicated by black circles and the radii denote the astrometric uncertainty (see Table~\ref{tab:coords}). The  sizes  of the 7~mm radio source or supernebula  \citep{turner04} and compact CO cloud D1 \citep{turner17} are shown as green and mauve ellipses respectively. The small circles at the centers of these ellipses indicate the central positions of the sources and the radii of the circles represent the astrometric uncertainty as given in Table~\ref{tab:coords}. 
The position of the 1.3~cm source (Secondary-East) of  \citet{turner04} is shown in cyan. The white circles indicate the FWHM spatial resolution of the images with the radii indicated. For (b), the radii are given for the F110W, F160W and F222M filters respectively.
}
\label{fig-astrom}
\end{figure*}

In Fig.~\ref{fig-astrom}, we show the ACS/HRC F435W, F550M, F814W image and the NICMOS/NIC2 F110W, F160W, F222M image of the central region of NGC 5253 mapped onto the Gaia DR2 reference frame. The NICMOS image is oversampled with a pixel size of 0\arcsec.04 (Table~\ref{tab:images})
The two clusters (\#5 and \#11), the supernebula position and size, and the CO cloud D1 position and size (\citep{turner04, turner17} are shown. It is immediately obvious from the HRC image in Fig.~\ref{fig-astrom}(a) that cluster \#11 is not coincident with the supernebula or the CO cloud D1. Both of these radio sources lie in a highly obscured region between clusters \#5 and \#11. We thus find that there are 3 clusters at the center of NGC 5253. There appears to be no cluster coincident with the 1.3 cm radio source.

The sources visible in the NICMOS image Fig.~\ref{fig-astrom}(b) are wavelength dependent. Cluster \#5 is bright in the F110W filter, which contains P$\beta$ nebular emission. The supernebula or embedded cluster is apparent at F160W and dominates in the F222M band where it is unresolved with a radius $<1.6$~pc.
Cluster \#11 is clearly present in the F110W and F160W images but is much fainter in F222M compared to the supernebula and is not resolved from this source.
We thus recognize 3 different sources at the NICMOS wavelengths. 
Most previous work has associated cluster \#11 with the supernebula. This explains the finding by \citet{calzetti15} that cluster \#11 is a factor of 2--2.5 brighter in the F110W and F160W bands than predicted by their SED fits. They attribute this excess flux to hot dust emission associated with \#11. It is now clear that the supernebula is contributing in the NICMOS bands and becomes the dominant source in the F222M band. \citet{cohen18} obtained K-band imaging at 0\arcsec.1 resolution with Keck II and associate the hot dust emission with the supernebula rather than cluster \#11.

\subsection{Sizes of the \hst clusters}\label{sizes}

The sizes of the two \hst clusters (\#5 and \#11) have not been measured before. To do this, we use similar methods to those in \citet{ryon15, ryon17}. We characterize the two-dimensional light profiles of the clusters  with the fitting package GALFIT \citep{peng02,peng10}.  We use the ACS/HRC images because they have the highest resolution (native pixel scale of $0\arcsec.025$) and are thus most likely to resolve the clusters if they are extended relative to the point spread function (PSF). The F814W filter was chosen because it is least likely to be affected by dust, though it does contain nebular emission. We also obtained an F814W image from \citet{calzetti15} that has had the nebular emission subtracted. 

GALFIT produces a model fit by convolving a model image with a provided PSF and comparing the result to
the observed data. We create a stellar PSF from several bright, isolated stars in the F814W image by using \verb+pstselect+ and \verb+psf+ within DAOPHOT in IRAF. We oversample the empirical PSFs by a factor of 10. 

\begin{deluxetable*}{lllcc}
\tablecolumns{5}
\tablecaption{Revised Properties of the Central SSCs in NGC~5253\label{tab:props}}
\tablehead{
\colhead{Parameter} & \colhead{Cluster \#5}  & \colhead{Cluster \#11}  &  \colhead{Supernebula}
& \colhead{References} \\
}
\startdata
Age (Myr) & $1\pm1$ & $1\pm1$ & $ < 1$ & 2,2,1\\
Mass (\Msun) & $7.5\pm0.3 \times 10^4$ & $2.5\pm0.6 \times 10^5$ & $2.5\pm0.9 \times 10^5$ & 2,2,3\\
Radius (pc) & $<0.6$ & $\la0.6$ & $0.80 \times 0.32$ & 1,1,4\\
A$_{\rm V}$ (mag) & 1.4 (foreground) & 50 (mixed) & embedded & 2,2,4\\
Projected separation (mas, pc) & 362, 5.82 & 94.2, 1.52 &  0, 0 & 1,1,1\\
\enddata
\tablerefs{1. This paper, 2. \citet{calzetti15}, 3. \citet{turner17} , 4. \citet{turner04}}
\end{deluxetable*}

The region containing clusters \#5 and \#11 is complex. 
With GALFIT, we attempt a range of models including single and multiple components. These components consist of the EFF (or Moffat) light profile, which has been shown to describe the light profiles of resolved young star clusters very well \citep*[e.g.][]{elson87, mackey03}, and/or the empirical PSF to represent the clusters of interest and other features within 20 to 30 pixels. Many GALFIT runs failed or did not converge, likely due to the complex nature of the region. 
Overall, we find that cluster \#5 is indistinguishable from the PSF in either image and thus is unresolved. The ACS/HRC F814W stellar PSF has a FWHM of 2.9 $\pm$ 0.1 pixels, which corresponds to an upper limit to the half-light radius of R$_{\mathrm{eff}} <\mathrm{FWHM}/2 <1.45$~pixels, or $<0.58 \pm0.05$~pc at the distance of NGC~5253. Cluster \#11 appears more elongated than the stellar PSF, but it suffers from higher and more complex extinction than cluster \#5 (Sec~\ref{opt}; \citealt{calzetti15}). The few EFF profile fits to cluster \#11 that converged did not appear to match the actual structure of the cluster or had unphysical effective radii. Therefore, we cannot reliably estimate the size of cluster \#11 using GALFIT. Because it appears close in size to the stellar PSF, we will assume it is $<0.6$~pc.

\section{Discussion and Conclusions}\label{disc}

We have established through re-mapping the \hst images to the Gaia reference frame that there are 3 SSCs at the center of NGC 5253. The properties of these clusters are summarized in Table~\ref{tab:props}. The age of the central starburst is often given in the literature as 3--5\,Myr, based on the presence of broad Wolf-Rayet (W-R) stellar emission features in the optical spectrum of the central region of NGC~5253
\citep{schaerer97b, monreal10}. \citet{calzetti15} derived much younger ages for clusters \#5 and \#11 of $1\pm1$\,Myr using 13 band \hst photometry. These ages appeared to contradict the presence of W-R features but \citet{smith16} showed that the W-R features in the UV and optical spectra of cluster \#5 arise from hydrogen-rich very massive stars with masses $>100$ \Msun\ at an age of 1--2 Myr. Given that \#5 and \#11 are $1\pm1$ Myr old, it seems likely that the supernebula is even younger. The supernebula is still embedded in its natal material with molecular CO (3-2) gas in the form of cloud D1 \citep{turner17}. However, clusters \#5 and \#11 have no detected CO (3-2), suggesting that they are older. Cluster \#11 in particular is noteworthy in having no detected CO (3-2) material, given that it is has a large dust component which is mixed in with the stars \citep{calzetti15}.
We thus suggest in Table~\ref{tab:props} that the age of the supernebula is probably $<1$~Myr and the embedded cluster is still potentially forming.

Turning now to the cluster masses, the masses for clusters \#5 and \#11 are from the SED fitting of \citet{calzetti15}. 
We note that although we find that the supernebula is contributing to the IR photometry measured by \citet{calzetti15} for \#11, the mass derived by these authors should still be accurate because they could not fit the excess IR flux. They find that \#11 is anomalously bright by a factor of 2--2.5 at F110W and F160W than predicted by their best fitting models covering the FUV to the IR. The mass for the supernebula in Table~\ref{tab:props} is the virial mass measured from the width of the CO(3-2) emission from \citet{turner17}. The masses of cluster \#11 and the supernebula are remarkably similar and may suggest that this is the maximum cluster mass that can be formed in the conditions of the central environment.

Clusters \#5 and \#11 are unresolved and have upper limits of 0.6~pc for their half-light radii. The supernebula is resolved \citep{turner04} but this size refers to the ultracompact H\two\ region and not the central ionizing cluster. The upper limits to the radii are in accord with those expected for very young massive clusters, which are usually very compact. In a study of the star clusters in M51, \citet{chandar16} find that the most compact clusters are $<10$~Myr old with typical half-light radii of 1.3~pc for masses $> 6 \times 10^4$~\Msun. \citet{leroy18} find FWHM sizes of 2--3~pc for knots of dust emission associated with 14 forming SSCs in the nuclear starburst of NGC~253 from ALMA observations.

Overall, cluster \#11 appears to be a twin of the supernebula in terms of the mass and high dust content. They have formed very close together with a small projected separation of 1.52 pc. Cluster \#11 appears to be slightly older because it is not embedded and is visible as a cluster at UV and optical wavelengths. It, moreover, appears to have no detectable CO(3-2) unlike the supernebula.

Finally, we consider the fate of the three nuclear SSCs at the center of NGC 5253. Nuclear star clusters are found in at least 70\% of all galaxies \citep{boker02}. Two main formation mechanisms are usually considered: in-situ formation at the galactic center from infalling gas or migration of stellar clusters to the center \citep*[e.g.][]{antonini15}. In NGC~5253, we are clearly witnessing the formation of massive clusters at the center from infalling gas. Given that they are in the potential well of the galaxy, it seems likely that they will merge, given their close proximity and similar radial velocities \citep{cohen18}. These very young SSCs may therefore be the building blocks for nuclear star clusters.

\acknowledgments
We thank Sara Beck, Michelle Consiglio and Jean Turner for very useful and illuminating discussions on their radio data at the 23rd Guillermo Haro Workshop. We also thank Daniela Calzetti for providing us with emission line subtracted \hst images. We thank the referee for their constructive comments.

Based on observations made with the NASA/ESA Hubble Space Telescope, obtained from the Mikulski Archive for Space Telescopes (MAST) at the Space Telescope Science Institute, which is operated by the Association of Universities for Research in Astronomy, Inc., under NASA contract NAS5-26555. 

This work has made use of data from the European Space Agency (ESA)
mission {\it Gaia} (\url{https://www.cosmos.esa.int/gaia}), processed by
the {\it Gaia} Data Processing and Analysis Consortium (DPAC,
\url{https://www.cosmos.esa.int/web/gaia/dpac/consortium}). Funding
for the DPAC has been provided by national institutions, in particular
the institutions participating in the {\it Gaia} Multilateral Agreement.

The HST data used in this analysis are available in the MAST archive at \dataset[10.17909/t9-8d2d-4k92]{https://doi.org/10.17909/t9-8d2d-4k92}. The images registered to the Gaia DR2 frame are available as a High Level Science Product at \dataset[10.17909/t9-75q2-x997]{https://doi.org/10.17909/t9-75q2-x997}.

\facility{HST (ACS, NICMOS)}
\software{DrizzlePac \citep{gonzaga12}, GALFIT \citep{peng02,peng10}, DAOPHOT \citep{stetson87}}

\bibliographystyle{aasjournal}
\bibliography{references}

\end{document}